\documentclass[conference,10pt]{IEEEtran}
\usepackage[T1]{fontenc}
\usepackage{diagbox}
\usepackage{makecell}
\usepackage{floatrow}
\usepackage{url} 
\usepackage{graphicx}
\usepackage{subfig}
\usepackage{amssymb}
\usepackage{amsbsy}
\usepackage{mathrsfs}
\usepackage{mdwlist}
\usepackage{float}
\floatstyle{plaintop}
\restylefloat{table}

\newcommand{\figref}[1]{Figure~\ref{#1}}

\begin{document}

\title{HTTP adaptive streaming with indoors-outdoors detection in mobile networks}
\author{
\IEEEauthorblockN{Sami Mekki, Theodoros Karagkioules, and Stefan Valentin}
\IEEEauthorblockA{Mathematical and Algorithmic Sciences Lab,\\
France Research Center, Huawei Technologies France SASU\\
\{sami.mekki, theodoros.karagkioules, stefan.valentin\}@huawei.com}
}

\maketitle

\begin{abstract}
In mobile networks, users may lose coverage when entering a building due to the high signal attenuation at windows and walls. Under such conditions, services with minimum bit-rate requirements, such as video streaming, often show poor Quality-of-Experience (QoE). We will present a Bayesian detector that combines measurements from two Smartphone sensors to decide if a user is inside a building or not. Based on this coverage classification, we will propose an HTTP adaptive streaming (HAS) algorithm to increase playback stability at a high average bitrate. Measurements in a typical office building show high accuracy for the presented detector and superior QoE for the proposed HAS algorithm.
\end{abstract}
\section{Introduction}
\label{sec:Intro}
In 2016, video streaming generated 60\% of the global IP traffic in mobile networks \cite{cisco-2017}.
The lion's share of this traffic was produced by HTTP Adaptive Streaming (HAS) technologies \cite{Sandvine-US}, according to the Dynamic Adaptive Streaming over HTTP (DASH) standard \cite{DASH} or the HTTP live streaming (HLS) draft \cite{pantos11:hls_ietf}. With HAS, a video stream is encoded in multiple qualities and segmented into a number of small files. Each of these so-called segments or chunks contains an approximately constant duration of video and is downloaded by the client via the HTTP protocol. The client employs a bitrate adaptation algorithm, called HAS policy, that controls quality and request time according to the filling level of its local playback buffer and according to the measured throughput. 

Although HAS policies are designed to cope with some throughput variation they often fail in mobile networks \cite{7511206,piStream}. Depending on the current user location, mobile throughput may change drastically as a result of physical effects during radio propagation. While the client's playback buffer typically protects the stream from short channel outages caused by Fast Fading, slow effects such as Path Loss and Shadowing can cause much longer throughput decreases. Being unaware of the state of the wireless channel, most HAS policies may react too late and not aggressive enough to compensate for such long degradations of wireless throughput. This is a common case if a user is obstructed by a large obstacle or moves inside a tunnel or a building \cite{7511206}. The target of this paper, is to provide a fast and aggressive HAS decision if (and only if) the user's current mobile coverage requires it.

\subsection{Related work}
HAS policies can be classified according to their input dynamic. Throughput-based policies adapt according to measured TCP throughput, which requires a sufficient number of probes in order to obtain a reliable observation \cite{Panda}. On the other hand, buffer-based policies \cite{Netflix,BOLA} adapt the requested bitrate according to the client's playback buffer. 

Since it takes a long time until a coverage loss affects application-layer buffers and TCP throughput, the mobile streaming community started to include information from lower protocol layers into HAS policies. While \cite{piStream} obtains the link-layer throughput directly from the LTE scheduling grant, the authors of \cite{bao15:has_chstate,7511206} decided to adapt to the received signal power. Although it cannot be directly translated into throughput, signal power provides an accurate measurement of the current coverage situation and is readily available at the application layer of modern Smartphones without additional battery consumption.

\subsection{Main contributions}
Following the idea of characterizing a user's coverage situation by signal power, we go a step further. In addition to the signal power received from the cellular modem, we use information from the Smartphone's localization sensors, e.g., its Global Navigation Satellite System (GNSS) receiver, to indicate a coverage loss. We will provide strong experimental evidence that this combination highly increases the estimation accuracy compared to using cellular signal power alone. 

Briefly, the contributions of this work are as follows:
\begin{itemize}
\item \emph{Indoors-outdoors detector:} In Section \ref{IOAlgo}, we present a Bayesian detector that combines signal measurements from the cellular and GNSS receiver. The result is a binary estimate (i.e., indoors or outdoors) of the user's coverage situation in real time.
\item \emph{Indoors-outdoors-aware HAS policy:} In Section \ref{AdaptiveStreamingsection}, we apply the indoors-outdoors detector described in Section \ref{IOAlgo} to improve the performance and stability of a classic adaptive streaming policy \cite{Netflix}.
\item \emph{Experimental verification:} In Section \ref{resultssection}, we present the setup and results of our measurement campaign.\footnote{A video illustrating the measurement scenario and the QoE gain for adaptive streaming is available at \url{https://youtu.be/qJibE2N37Yk}.}. The results consistently show substantial gains in video bitrate, fluency and stability, compared to the baseline policy \cite{Netflix}.
\end{itemize}
These results demonstrate that accounting for the coverage \emph{context} of a user is a promising approach for designing adaptive streaming policies. 

\section{System model}
According to the DASH standard \cite{DASH}, a streaming client employs an HAS algorithm to choose a video quality. For this quality, the client progressively downloads each segment until the maximum buffer occupancy ($B_{max}$) is reached. The QoE eventually depends on the filling level of this playback buffer, which is defined by the throughput of the radio link. 

Based on measurements of the received signal power, we compute the throughput according to Shannon's equation as follows. Achievable rate at time index $i$ is given by
\begin{eqnarray}
 {C}_i= B_i  \log_2(1+\gamma_i),
\label{eq:ShannonLaw}
\end{eqnarray}
where $\gamma_i$ is the signal to interference noise ratio (SINR) and $B_i$ is the user effective bandwidth which corresponds to $90\%$ of the total bandwidth in the 4G standard \cite{UMTS-LTE-Sesia-2009}. We assume that $\gamma_i  = P_i/\theta_i$ where $P_i$ is the received signal power measured over bandwidth $\omega_{s}$  and $\theta_i$ is the sensitivity threshold, above which the receiver is capable of detecting and processing the received signal. We truncate the value of $\gamma_i$ at $20$dB since the highest modulation scheme is obtained at SINR$>=20$dB.

The received signal power is available at the application layer of most mobile devices \cite{Android-signal-api-2016}. Without loss of generality, we assume that interference is included in the sensitivity threshold. Since the studied mobile devices may associate to a different mobile network type any time, thresholds for 2G, 3G and 4G have to be considered. We summarize all used parameters in Table \ref{tab:ThrSensitivitySys}. Details are provided in the standards \cite{GSMTS:GSM.05.05.Sec.6.2-GSMthr,3gpp:TS.25.101.Sec.7.3.1-3Gthr,3gpp:TS.36.521.Sec.7.3.1-LTEthr} for 2G, 3G and 4G, respectively.
\begin{table}[h]
\resizebox{\columnwidth}{!}{
\centering
\caption{Parameters for the studied mobile networks}
\label{tab:ThrSensitivitySys}
\begin{tabular}{|l||c|c|c|} \hline 
Parameters 		& 2G & 3G & 4G\\ \hline\hline
Received power $P_i$ & RSSI & RSCP & RSRP  \\\hline
Measured bandwidth $\omega_s$& $200$kHz & $3.84$MHz & $15$kHz	\\ \hline
Data bandwidth $B_i$ & $200$kHz & $5$MHz & $18$MHz   \\ \hline
Sensitivity threshold $\theta_i$ [dBm/$\omega_s$]	&  -104   &  -106 & -94 \\ \hline
\end{tabular}}
\end{table}

\begin{figure}[t]
\includegraphics[width=0.8\linewidth]{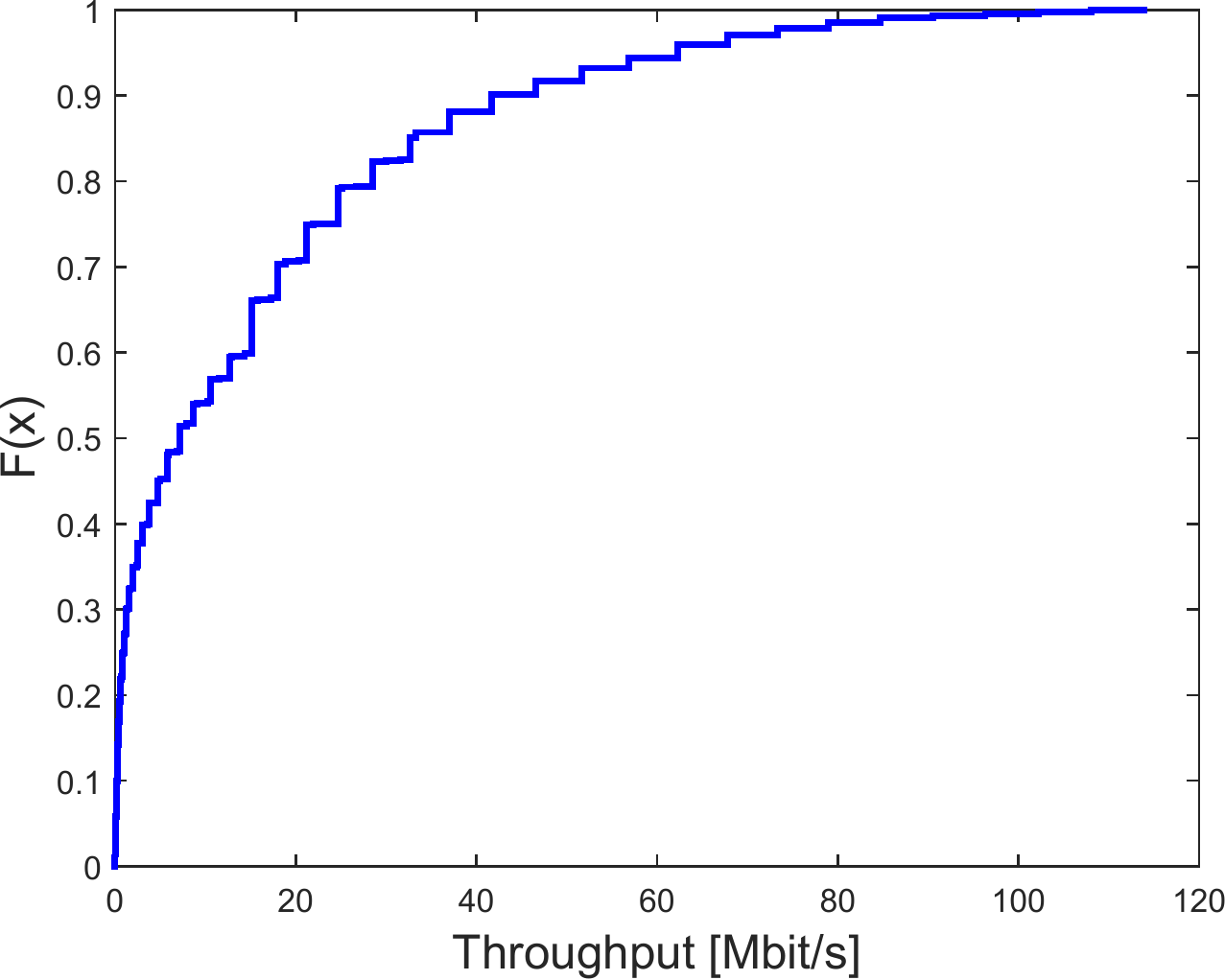}
\caption{Empirical CDF of the cell throughput in the downlink}
\label{fig:TotaleCDF}
\end{figure}

Using the obtained SINR, the cell throughput is then computed over the complete bandwidth of the respective network type as in (\ref{eq:ShannonLaw}). Computing this quantity for the complete data set provides the empirical CDF in \figref{fig:TotaleCDF}.

To obtain the throughput for a single user, we assume a round-robin scheduling policy where all users are served equally. Thus, the total cell throughput is equally shared among $K$ users as $ {C}_{i,k}= {C}_i/K$ where $k=1,\ldots,K$. Note that we have to limit our study to $K\leq8$, since this is the maximum number of users per channel in 2G \cite{GSMTS:GSM.05.01}.

The received signal power $P_i$ was measured over several days inside and outside of a typical multi-floor office building in downtown Paris, France. Off-the-shelf mobile devices were used, namely five HUAWEI Mate 8 Smartphones running Android 6.0 Marshmallow \cite{web-android-6} with different network providers. The Smartphones automatically selected all transmission (e.g., Physical-layer rate) and associated to 2G, 3G, or 4G networks over the course of the measurements. All data was recorded at the application of these devices without direct access to the mobile networks. In addition to signal power, localization data and ground truth about the user's position (indoors or outdoors) were recorded.

\section{Indoors-Outdoors detection}
\label{IOAlgo}
In this section, we present our Bayesian classifier for a user's coverage situation. Our algorithm detects if a user is indoors or outdoors, based on the a posteriori information from Smartphone sensors. The first input is the received signal power as provided by the cellular modem according to \cite{ETSI-TS-36.214}. The second input is the confidence radius of the location, provided by the active localization sensor, e.g., the Smartphone's global positioning system (GPS) chip set. In the following, we will discuss the statistics of both input variables and describe a method to combine them for indoors-outdoors detection.

\subsection{Analysis of received signal power}
\label{subsec:RSSIanalysisIOdetect}
\begin{figure}
\includegraphics[width=0.8\linewidth]{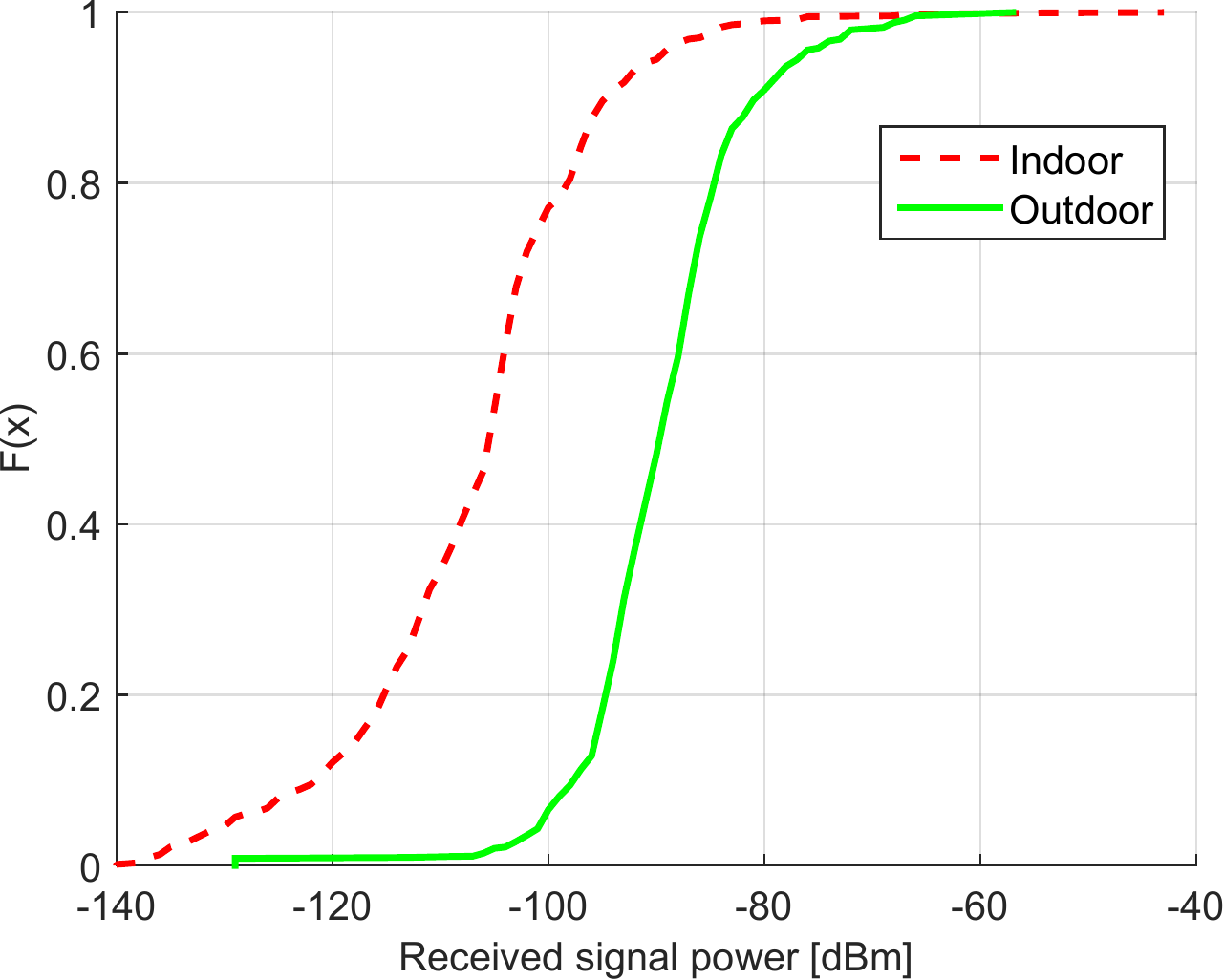}
\caption{Empirical CDF for measured received signal power}
\label{fig:RSSIeCDF}
\end{figure}

\figref{fig:RSSIeCDF} shows the empirical cumulative distribution function (CDF) of the received signal power measurement, separated by the recorded ground truth (i.e., indoors or outdoors). The significant offset between the two CDFs results from substantial differences in radio propagation, namely reflection, diffraction, and attenuation \cite{Goldsmith-model-2005}. Based on these statistics, we conclude that received signal power can be used to classify the user's coverage situation (i.e., either indoors or outdoors). We will proceed using the empirical probability density functions (pdf) of the received power, obtained by differentiating previously recorded CDFs.

\subsection{Analysis of confidence radius} 
To improve the detection performance, we propose the use of the confidence radius as provided by the localization system of mobile devices, e.g., \cite{Android-gps-2016}. It is important to highlight that confidence radios is the sole localization parameter used by our algorithm. Other parameters such as user position, speed and altitude are not required.

\begin{figure}
\includegraphics[width=0.8\linewidth]{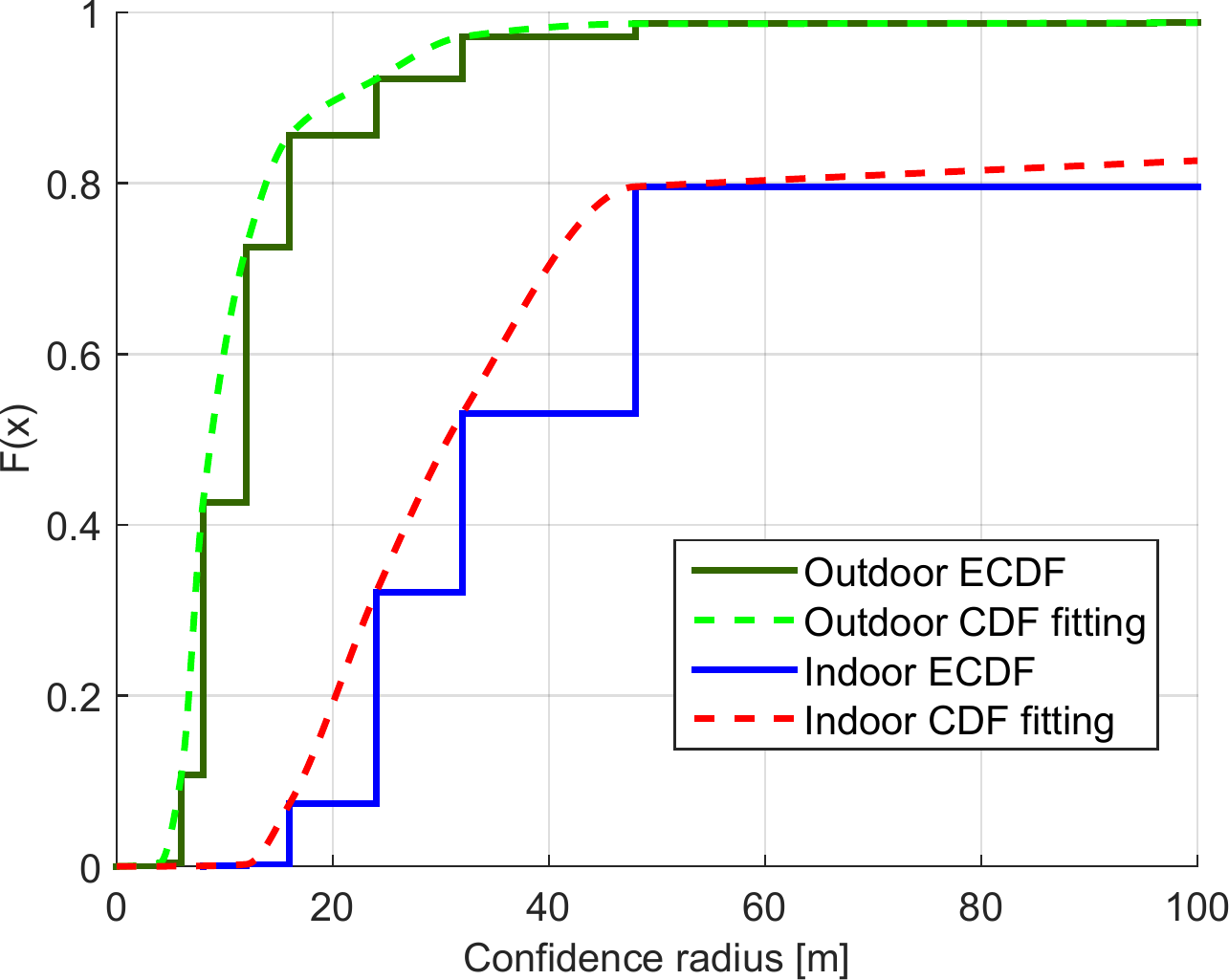}
\caption{Empirical CDF for measured confidence radius}
\label{fig:GPSaccCDF}
\end{figure}

The empirical analysis of the confidence radius shows sparse data for indoors and outdoors measurements as in \figref{fig:GPSaccCDF}. Such quantization is common in GPS chip sets and can be overcome by curve fitting, as shown by the dashed lines. The resulting curves will be used in the following algorithm


\subsection{Data fusion and detection}
The indoors-outdoors detection is based on maximum \emph{a posteriori} (MAP) estimation, where the unknown \emph{status} maximizes the a posteriori probability under a given observation. The input from the two sensors is combined by computing a joint a posteriori probability based on the distributions of (i) the received signal power and (ii) the confidence radius.

We denote by ${y}_\ell$ and $z_\ell$ the received signal power measure and the confidence radius value, respectively,  at time index $\ell$. The a posteriori probability is given by $APP(S_\ell)=P(S_\ell|y_\ell,z_\ell )$, where $S_\ell \in\{Indoor, Outdoor\}$. Following Bayes' law, the a posteriori probability becomes 
\begin{eqnarray} 
\label{eq:APPgpsRSSI}
APP(S_\ell)=\frac{P(S_\ell ,y_\ell ,z_\ell)}{P(y_\ell ,z_\ell)}=\frac{P(y_\ell ,z_\ell|S_\ell)P(S_\ell)}{P(y_\ell ,z_\ell)}.
\end{eqnarray}
Unlike for received power and confidence radius, no a priori information is available on $S_\ell$ and the two input variables are obtained from physically independent sensors. Consequently, the a posteriori probability is proportional to the product
\begin{eqnarray} 
\label{eq:APPgpsRSSI2}
APP(S_\ell) \propto P(y_\ell ,z_\ell|S_\ell)=P(y_\ell|S_\ell)P(z_\ell|S_\ell) \label{eq:proptoAPP}
\end{eqnarray} 
of the conditioned measurements. Here, $P(y_\ell|S_\ell)$ is the direct observation based on received signal power and $P(z_\ell|S_\ell)$ is the direct observation based on the confidence radius. Finally, the a posteriori probability can be expressed simply as
\begin{eqnarray}
APP(S_\ell) \propto  obs_y(S_\ell)~obs_{z}(S_\ell),
\label{eq:RSSIgpsAPPdetect}
\end{eqnarray}
where  $obs_y(S_\ell)=P(y_\ell|S_\ell)=\frac{p(y_\ell|S_\ell)}{\sum_{S_\ell} p(y_\ell|S_\ell)} $ and $obs_z(S_\ell)=P(z_\ell|S_\ell)=\frac{p(z_\ell|S_\ell)}{\sum_{S_\ell} p(z_\ell|S_\ell)}$ are the observations computed 
from received power and location measurements, as described previously.

The coverage classification (i.e., indoors or outdoors) is then given by the most probable value of $APP(S_\ell)$, i.e., the value maximizing the a posteriori probability.

\section{Adaptive streaming}
\label{AdaptiveStreamingsection}
In this section we will describe an HAS policy based on coverage classification.

\subsection{Adaptation based on buffer occupancy and segment size}
\begin{center}
\begin{figure}[t]
\includegraphics[width=0.75\columnwidth]{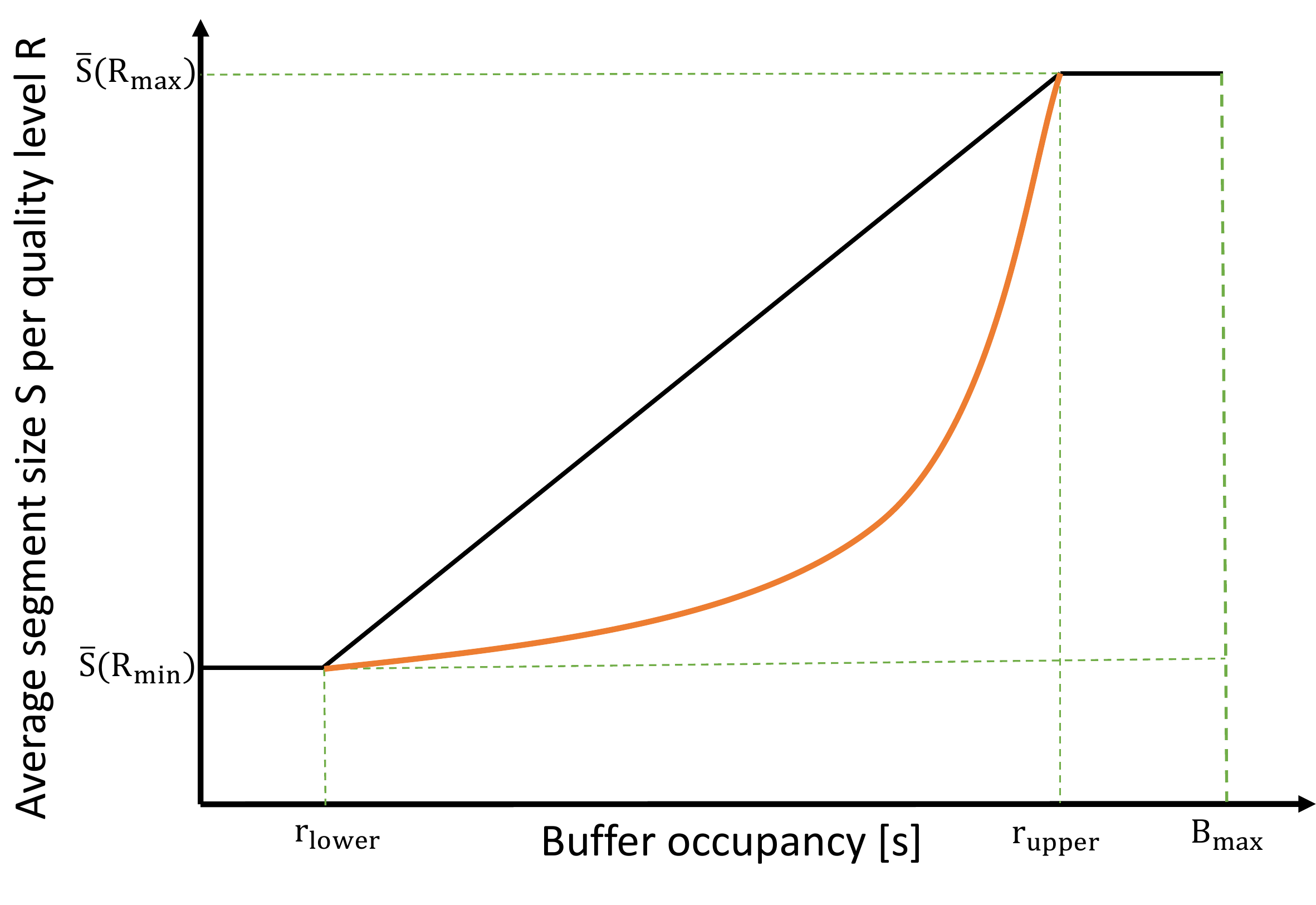}
\caption{Segment map of the baseline (black line) and of the proposed IOBBA policy (red line)}
\label{fig:chunkmap}
\end{figure}
\end{center}

A common buffer-based HAS policy is described in \cite[Section VII]{Netflix}. This policy is employs a segment map which is defined in the space $[0,B_{max}]$ and $ [\bar{S}(R_{min}),\bar{S}(R_{max})]$, where $\bar{S}$ is the mean segment size of representation $R_i$, $i \in(1,2\dots n)$, as depicted by the black line in \figref{fig:chunkmap}. The map is defined by two thresholds: i) an upper threshold $r_{upper}$ that drives the policy to select the maximum quality available ($R_{max}$) once the instantaneous buffer occupancy $B(t_j)$, at time $t$ at the end of the download of the $j^{th}$  segment, $j \in\{1,2\dots J\}$, surpasses it and ii) a lower threshold $r_{lower}$ that dictates the lowest available quality ($R_{min}$) if $B(t_j)<r_{lower}$. In the buffer region $r_{lower}<B(t_j)<r_{upper}$, the policy may use any non-decreasing function to select the quality of the next segment to be downloaded \cite{Netflix}. We will refer to this adaptation principle as the baseline for the remainder of the paper.

\subsection{Indoors-Outdoors aware HTTP Adaptive Streaming}	
We present the HAS policy \emph{Indoors-Outdoors aware Buffer Based Adaptation} (IOBBA), based on coverage classification. IOBBA aims at minimizing the re-buffering events for mobile streaming users that move into an area with poor coverage, e.g., inside a building. To this end, the HAS policy has to react faster to throughput decreases in order to avoid buffer underruns. Moreover, the policy should increase the video bitrate conservatively, in order to avoid video rate oscillations at the edge of a poorly covered region. Both design principles should add a minimum penalty to the average video bitrate. Following these targets, we made two modifications to the baseline algorithm \cite[Section VII]{Netflix}. 

\subsubsection{Aggressive response to reduced buffer occupancy}
Instead of using a linear function for the buffer region $r_{lower}<B(t_j)<r_{upper}$, we employ the function $f(B(t_j)=\alpha \cdot \beta^{B(t_j)}$, where $\alpha$ and $\beta$ can be computed at the boundary values $f(r_{lower})$ and $f(r_{upper})$. This exponential function is illustrated by the red line in \figref{fig:chunkmap}. To increase stability and to reduce the probability of buffer starvation, we use a constant value for $r_{lower}$. This allows for a fast reaction to the immediate buffer depletion, which is a direct consequence of the throughput reduction, when moving into a building. This updated segment map is used when the user location is detected indoors. In this way, the user will experience smoother buffer depletion and thus a reduced probability of re-buffering events. To prevent further loss in the mean video bit-rate, a linear function is used instead, as soon as the user's location is detected outdoors.

\subsubsection{Conservative video bit-rate increase}
Indoors, the IOBBA policy increases video bitrate more conservatively than outdoors. While the user is considered to be indoors, video bitrate is only increased after $m$ sequential requests upgrade requests have been indicated by the segment map. With this principle, video bitrate oscillation becomes less likely.

\section{Measurement results}
\label{resultssection}
We experimentally evaluate various QoE factors for IOBBA. For each factor, we study the difference to ground truth and to a baseline algorithm from literature.

\subsection{Methodology}
To study the effect of estimation error on the QoE, we study \emph{IOBBA-True} where our HAS policy from Section \ref{AdaptiveStreamingsection} operates on the ground truth and \emph{IOBBA-Detected} where the result of our MAP estimator is used. During our one-week measurement campaign, we obtained 35 traces files for a variable number of mobile users $(k \in\{1,2,\dots, 8\})$. Furthermore, we investigated the maximum buffer occupancy $B_{max}$. This factor is important as it may substantially differ between different streaming services (e.g., tens of seconds for live streaming, hundreds of seconds for video-on-demand).

To parameterize our HAS policy, we computed the lower threshold dynamically calculated as in \cite{Netflix} when the user is outdoors. Indoors, we used the constant value $r_{lower}=0.3\times B_{max}$. The upper threshold is constant with $0.9\times B_{max}$ for all policies and the video rate increase threshold is $m=3$. 

As streaming content, we have chosen the open movie Big Buck Bunny (BBB) \cite{BigBuckBunnieMovie}, which is recommended in the measurement guidelines of the DASH Industry Forum \cite{Forum2014}. BBB is of $9$:$56$ minutes duration and high motion. We encoded the movie with H.264 at 24 frames per second, using MP4 containers at a segment duration of 4 seconds. Following the recommendations of \cite{ITEC}, we selecting the video bit-rate levels according to the quantiles of the total throughput CDF as in Table \ref{quantiles}.
\begin{table}[]
	\centering
	\resizebox{\columnwidth}{!}{
		\caption{Video content characteristics}
		\label{quantiles}
		\begin{tabular}{|c|c|c|c|}
			\hline
			\begin{tabular}[c]{@{}c@{}}Quality\\  index\end{tabular} & Resolution & \begin{tabular}[c]{@{}c@{}}Max\\ encoding rate (kbps)
			\end{tabular} & {Quantiles of \figref{fig:TotaleCDF}}  \\ \hline\hline
			1                                                        & 320$\times$240          & 129                                                     & 0.1                                                              \\ \hline
			2                                                        & 480$\times$360          & 378                                                      & 0.2                                                                \\ \hline
			3                                                        & 854$\times$480          & 578                                                      & 0.25                                                           \\ \hline
			4                                                        & 1280$\times$720         & 1536                                                      & 0.4                                                                   \\ \hline
			5                                                        & 1920$\times$1080        & 3993                                                  & 0.5                                                                   \\ \hline
	\end{tabular}}
\end{table}

As QoE factors we studied:
\begin{enumerate}
	\item \emph{Mean selected video rate}, averaged over the video duration.
	\item \emph{Re-buffering frequency}, which is the number of re-buffering events over the video duration and
	\item \emph{Adaptation frequency}, which accounts for the amount of  video rate changes over the video duration.
\end{enumerate}
For each metric, we present arithmetic means over all measurements and the standard error at a confidence level of $95\%$.

\subsection{Accuracy of indoors-outdoors detection}
The detection is performed by the MAP estimator defined in Section \ref{IOAlgo}, based on the pdfs from previous measurements.

The confusion matrix in Table \ref{tab:RSSIconfusionMatrix} shows the accuracy of this estimation when only the received power is used as an input.
\begin{table}[h]
\centering
\caption{Confusion matrix for indoors-outdoors detection based on received power}
\label{tab:RSSIconfusionMatrix}
\begin{tabular}{|l||c|c|} \hline 
\diaghead{\theadfont $P\left(Detected|True\right)$}%
{Detected($D$)}{True state\\($S$)}& Indoor & Outdoor\\ \hline\hline
Indoor &  {\bf 0.8256}  &  0.083 \\ \hline
Outdoor & 0.1744 &  {\bf 0.917}   \\ \hline
\end{tabular}
\end{table}

Including the confidence radius as a second input leads to the detection accuracy in Table  \ref{tab:RSSIgpsConfusionMatrix}.
\begin{table}[h]
\centering
\caption{Confusion matrix for indoors-outdoors detection based on data fusion}
\label{tab:RSSIgpsConfusionMatrix}
\begin{tabular}{|l||c|c|} \hline 
\diaghead{\theadfont $P\left(Detected|True\right)$}%
{Detected($D$)}{True state\\($S$)}& Indoor & Outdoor\\ \hline\hline
Indoor & {\bf 0.9496}  &  0.0911 \\ \hline
Outdoor & 0.0504 &  {\bf 0.9089}   \\ \hline
\end{tabular}
\end{table}

This clearly shows the benefit of including this second variable as a $12\%$ accuracy increase for the indoor case. Over all cases, the MAP estimation over both variables reaches an accuracy of $P_{io}=92.98\%$. 

\subsection{IOBBA performance}
\begin{figure*}[!t]
     \subfloat[Mean video bit-rate vs. $k$]{ %
    \includegraphics[ width=0.33\linewidth]{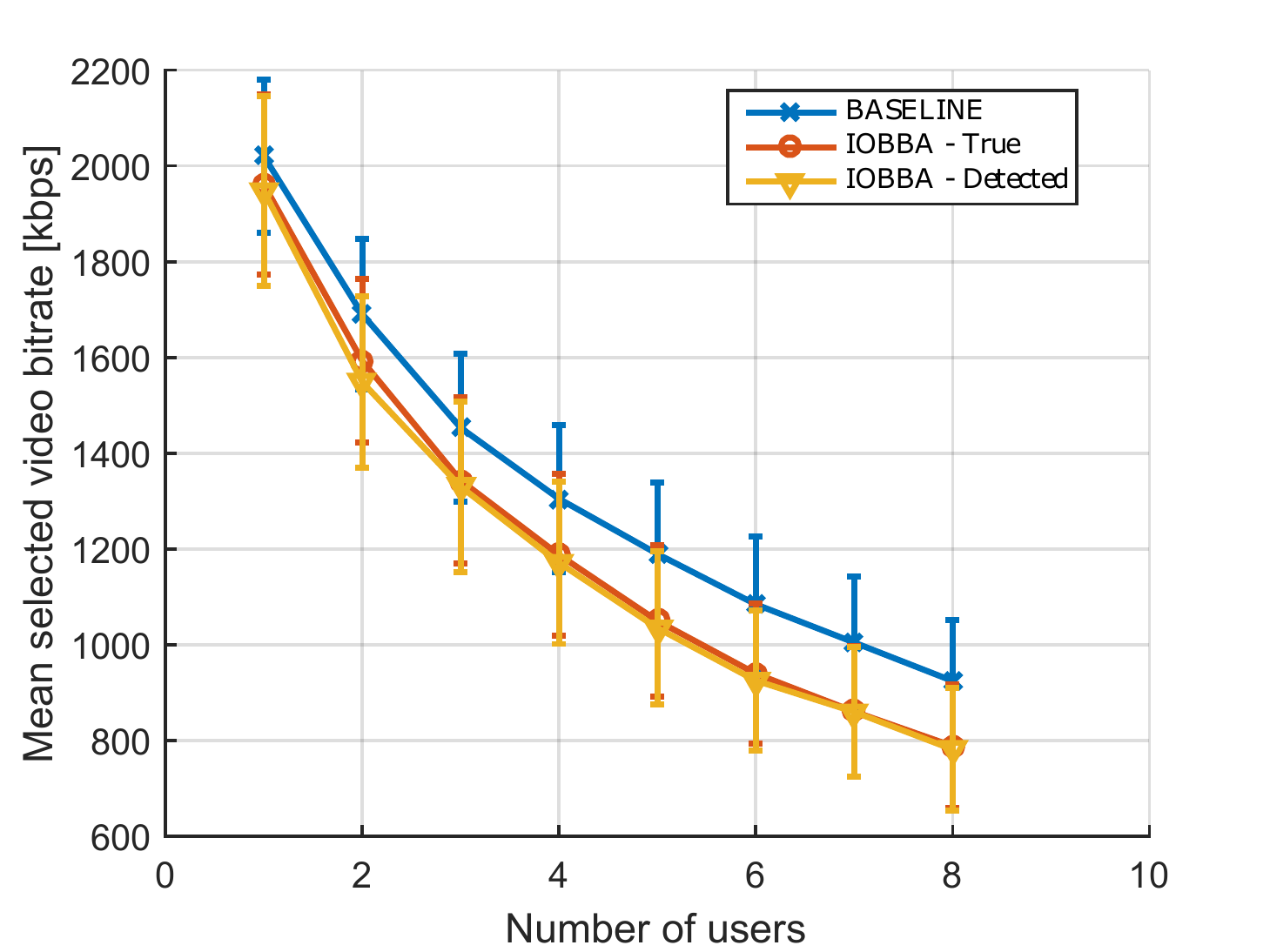} \label{fig:MeanVideoBitRate}
      }
     \subfloat[Adaptation frequency vs. $k$]{ %
       \includegraphics[ width=0.33\linewidth]{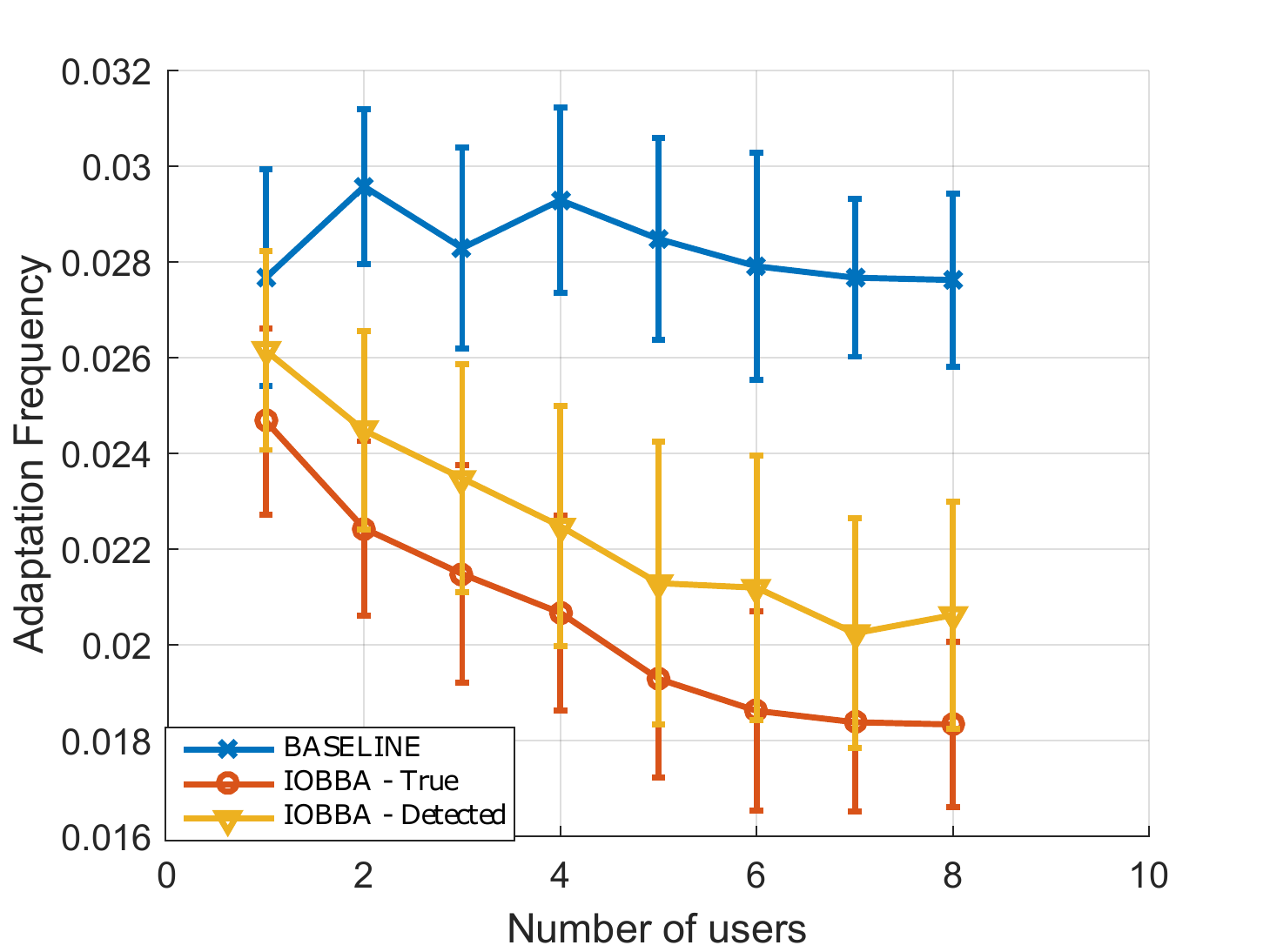}\label{fig:/AdaptationFrequency}
     }
     \subfloat[Re-buffering frequency vs. $k$]{ %
       \includegraphics[ width=0.33\linewidth]{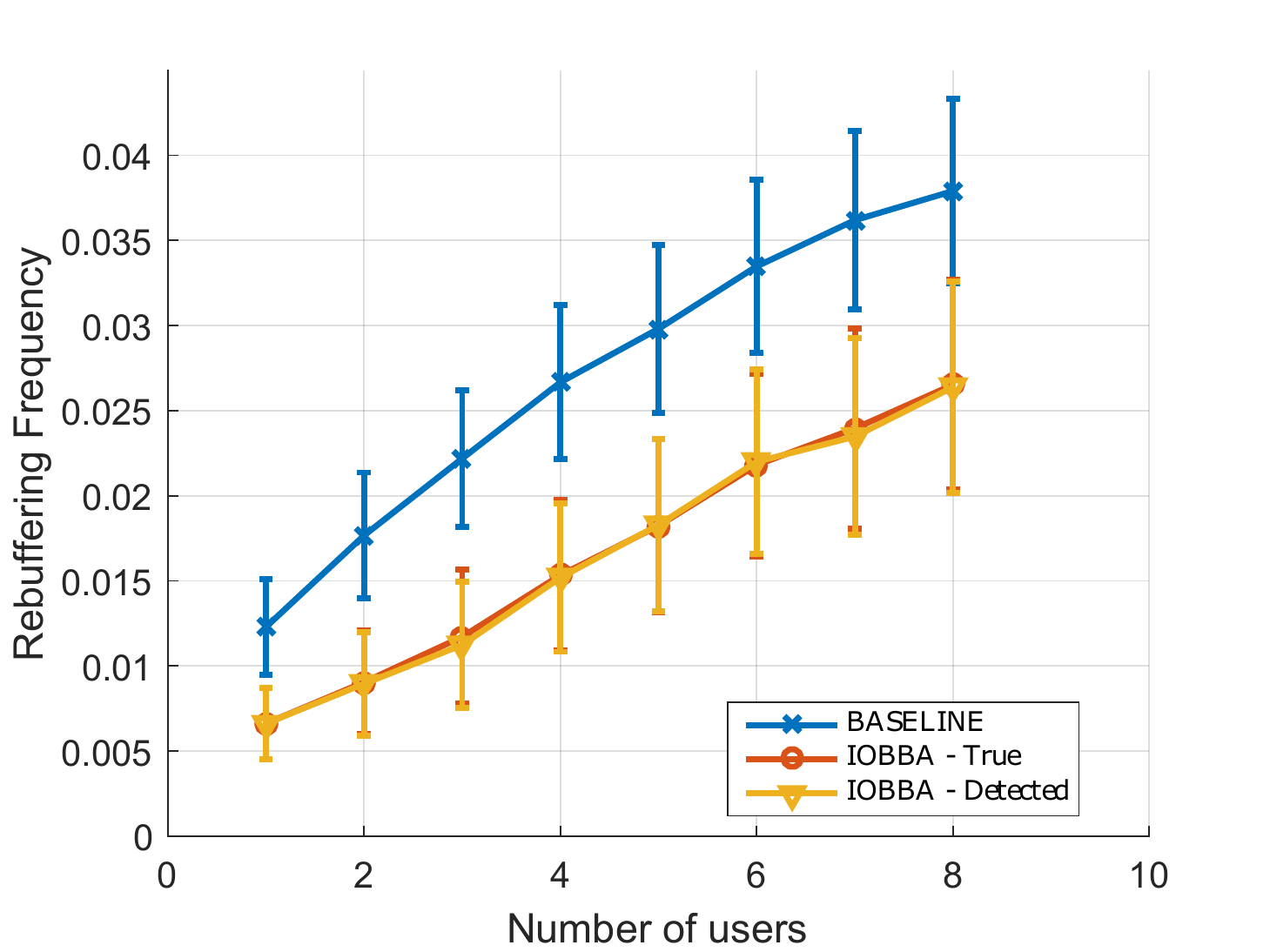}\label{fig:RebufferingFrequency}
     }
       \caption{ Experimental results for 95\% confidence intervals with $B_{max}=150$s}
     \label{fig:IOBBAresults}
\end{figure*}

\figref{fig:IOBBAresults} shows several QoE factors for an increasing number of users $k$. 
     \begin{figure*}[ht]
      \subfloat[Mean video bit-rate per $B_{max}$]{ %
    \includegraphics[ width=0.33\linewidth]{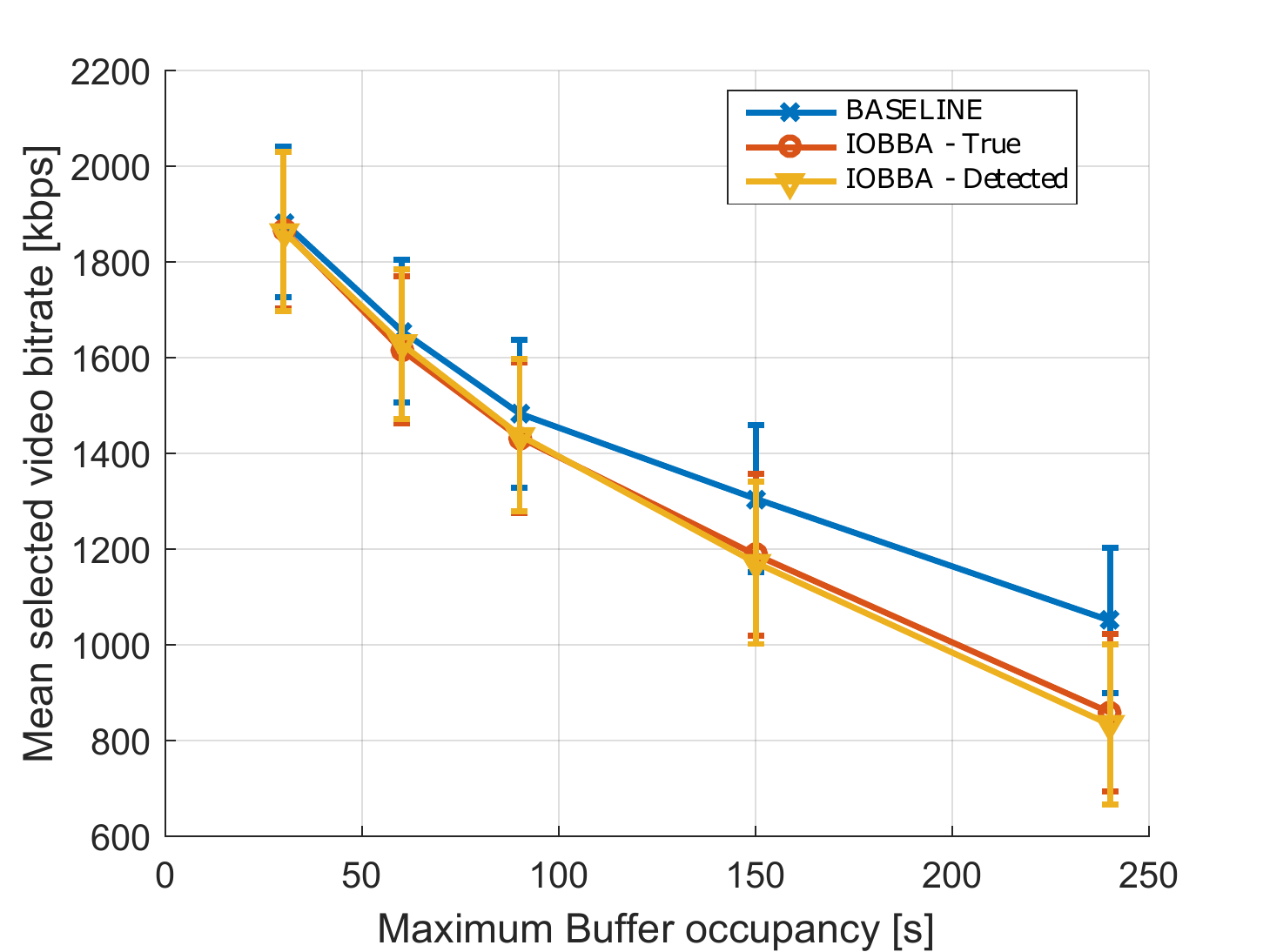}\label{fig:MeanVideoBitRate2}
      }
     \subfloat[Adaptation frequency per $B_{max}$]{ %
       \includegraphics[ width=0.33\linewidth]{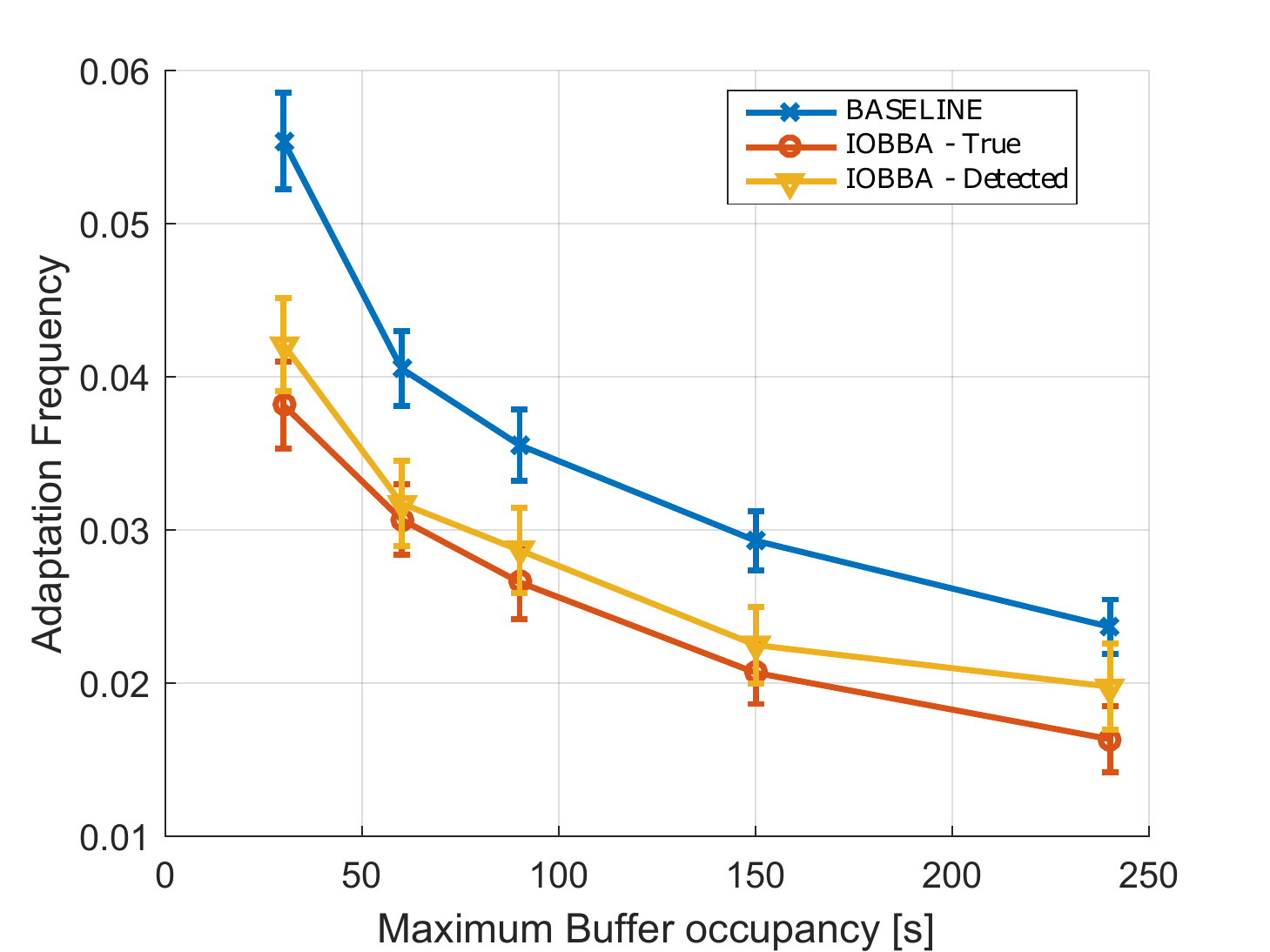}\label{fig:/AdaptationFrequency2}
     }
     \subfloat[Re buffering frequency per $B_{max}$]{ %
       \includegraphics[ width=0.33\linewidth]{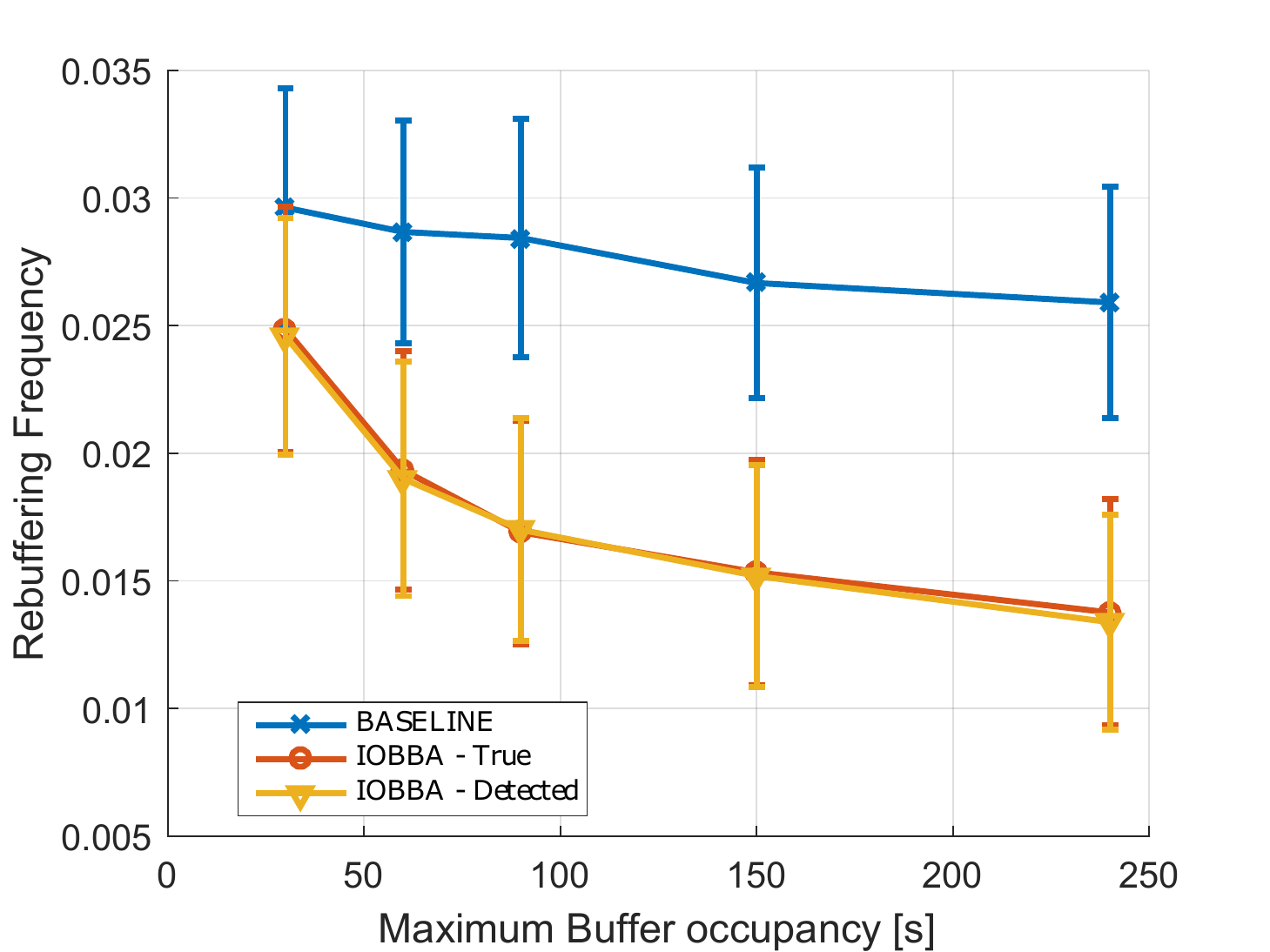}\label{fig:RebufferingFrequency2}
     }
     \caption{ Experimental results for 95\% confidence intervals with 4 users}
     \label{fig:IOBBAresults2}
\end{figure*}
As $k$ increases, the bandwidth-share assigned to each user is reduced and therefore the mean video bit-rate is diminished. Additionally, the frequency of re-buffering events is increased since the lower the throughput of each user the higher the probability of a buffer underrun. For the adaptation frequency, we notice that using indoor-outdoor detection increases the stability for increasing $k$. The baseline algorithm, however, suffers from an almost constant fluctuation of video quality even at high $k$. It is interesting that imperfect estimation shows a slight drawback for adaptation frequency but is insignificant for the other studied QoE factors. We can conclude that, compared to the baseline, \textit{IOBBA} improves the re-buffering frequency as well as adaptation frequency by $30\%$. This QoE gain comes only at a marginal cost for the mean video bit-rate. 

\figref{fig:IOBBAresults2} shows the studied QoE factors for an increasing target buffer level. A larger maximum buffer ensures that the video bit rates are sufficiently distant from each other in the segment map, thus reducing the probability of bitrate oscillation. Of course, the higher the maximum buffer occupancy, the higher the average buffer occupancy per stream, and therefore, the lower the re-buffering frequency.

We should mention that the mean video bit-rate is inversely proportional to the maximum buffer occupancy which is due the fact that the segment maps of larger buffers have a lower response rate to throughput increases. The larger the buffer the more time is spent to achieve $r_{upper}$ with segments of the lowest quality, which explains the reduction of IOBBA's mean bitrate in Figure \ref{fig:MeanVideoBitRate2}.

On the other hand, the smaller the maximum buffer occupancy, the smaller the distances between buffer values that indicate video bit-rate changes and, thus, the larger the adaptation frequency. This is why IOBBA outperforms the baseline by $20$ to $25\%$ in adaptation frequency and by $35$ to $40\%$ in terms of re-buffering. All in all, this is a substantial QoE gain at a reasonable penalty for the average bitrate.

\section{conclusion}
\label{concsection}
In this paper, we extended a common HTTP adaptive streaming (HAS) algorithm by a Bayesian coverage estimator. The resulting HAS policy, called \emph{IOBBA}, aggressively reduces the requested video quality if the user is estimated to enter an area of poor signal coverage (e.g., indoors). Then the policy stays conservative, until an area of high coverage is detected (e.g., outdoors). 

The coverage estimation requires no knowledge of user location and makes no use coverage or map data. The required input is entirely collected during run time, from two Smartphone sensors at the application layer. Estimation accuracy is in the order of 95\%, which is sufficient for high QoE gains.

Our experiments in a typical downtown office building show substantial improvements in video bit-rate, smoothness and stability, compared to a conventional HAS policy. This clearly demonstrates that making adaptive streaming aware of the user's current coverage situation can highly improve the QoE of mobile streaming.

\bibliography{IEEEabrv,InfoCom}
\bibliographystyle{IEEEtran}
\end{document}